# Roma versus Zaragoza[*]

J.L. Alonso[a], Ph. Boucaud[b], F. Lesmes[a] and A.J. van der Sijs[a]

[a]Departamento de Física Teórica, Universidad de Zaragoza,
Facultad de Ciencias, 50009 Zaragoza, Spain

[b]Laboratoire de Physique Théorique et Hautes Energies,
Université de Paris XI, 91405 Orsay Cedex, France[†]

The Roma and Zaragoza actions for chiral fermions on the lattice are shown to be essentially equivalent. The auxiliary fermion fields in the Roma model can be integrated out, and the resulting action is a special case of the Zaragoza approach.

We use this result to perform a mean-field study of the phase diagram of chiral Yukawa models in the Roma formulation.

## 1. INTRODUCTION

The Roma [1] and Zaragoza [2] proposals for chiral fermions on the lattice share a couple of important properties. Both formulations break the gauge symmetry explicitly, requiring gauge fixing and ghost fields. They preserve the global chiral symmetry, however, which restricts the possible counterterms. Furthermore, both models pass a useful test of their fermion content: a one-loop lattice calculation of the electroweak parameters $S$, $U$ and $T$ or $\Delta\rho$ correctly reproduces the continuum result [3,4]. This is to be contrasted with other models, where these parameters receive a contribution from non-decoupling doublers or supplementary fermions.

An apparent difference between the Roma and Zaragoza formulations is the way in which the fermion doublers are dealt with. In the Zaragoza philosophy the doublers are allowed to propagate freely but they are kept massless and non-interacting by suppressing their contribution to the vertices. This decoupling persists to all orders in perturbation theory. In the Roma approach, a non-interacting fermionic partner of opposite chirality is introduced for each physical fermion.

[*]Preprint Zaragoza DFTUZ/94/22, Orsay LPTHE 94–103, November 1994.
Contribution to Lattice '94 by A.J. van der Sijs (arjan@sol.unizar.es). Work supported by EC contracts CHRX-CT92-0051 and ERBCHBICT941067.
[†]Laboratoire associé au CNRS

This allows one to write down a Wilson term to decouple the doublers via the standard mechanism. The auxiliary fermions have no relevant interactions and decouple.

This difference is only superficial, though. As we shall discuss here, both approaches are essentially equivalent [5]. The auxiliary fermions in the Roma approach can be integrated out, leaving one with an action of the Zaragoza type.

This allows us to explore the phase diagram of chiral Yukawa models based on the Roma approach, using mean-field methods applied earlier for Zaragoza models [6] (see also ref. [5]).

## 2. ZARAGOZA

The action of a fermion-gauge-Higgs model in the Zaragoza formulation is [2]

$$S = \overline{\psi}K\psi + \overline{\psi}^{(1)}(U-K)\psi^{(1)} + y\,\overline{\psi}^{(1)}\Phi\psi^{(1)}, \quad (1)$$

restricting ourselves to the terms involving fermions. The kinetic term contains the naive lattice Dirac operator $K$, which in momentum space has the diagonal form $i\slashed{k}(p) = i\sum_\mu \gamma_\mu \sin p_\mu$. The field $\psi^{(1)}$ which enters in the interaction with the lattice gauge field $U = \sum_\mu \gamma_\mu (U_{L\mu}P_L + U_{R\mu}P_R)$ and the Higgs field $\Phi = \phi P_R + \phi^+ P_L$ is defined as

$$\psi^{(1)}(p) \equiv F(p)\psi(p). \quad (2)$$

The 'form factor' $F(p)$ is required to vanish at the doubler momenta and to satisfy $F(0) = 1$. It suppresses interactions of the fermion doublers. A convenient choice, from the point of view of numerical simulations, is the most local one,

$$F(p) = \prod_\mu \cos(p_\mu/2), \quad p_\mu \in (-\pi, \pi]. \qquad (3)$$

With this choice, $\psi^{(1)}$ in $x$-space is the average of the fields $\psi$ at the corners of a hypercube.

Because the interaction terms involve linear combinations of $\psi$ fields at different sites, the action (1) is not gauge invariant but does respect the global chiral symmetry.

After a change of variables $\psi(p) \to \psi(p)/F(p)$ the action becomes

$$S = \overline{\psi} F^{-1} K F^{-1} \psi + \overline{\psi}(U-K)\psi + y\,\overline{\psi}\Phi\psi. \qquad (4)$$

In this representation the interaction terms are of the usual form while the kinetic term is modified. The propagator in the broken phase ($m = y\langle\phi\rangle$) is

$$S(p) = \frac{-i\slashed{s} + mF^2}{s^2 + m^2 F^4}\, F^2, \qquad (5)$$

with $s^2(p) = \sum_\mu \sin^2 p_\mu$. The overall factor $F^2$ suppresses poles at the doubler momenta, and the other factors of $F^2$ suppress the doubler masses.

## 3. ROMA

In the Roma approach [1] the action is

$$\begin{aligned}
S &= \overline{\psi}_L U_L \psi_L + \overline{\chi}_R K \chi_R + \overline{\psi}_R U_R \psi_R + \overline{\chi}_L K \chi_L \\
&\quad + \overline{\psi}_L W \chi_R + \overline{\psi}_R W \chi_L + y\,\overline{\psi}_L \Phi \psi_R + h.c. \quad (6)
\end{aligned}$$

The physical fermions $\psi$ are coupled to the gauge fields $U_{L,R}$ and the Higgs field $\Phi$. There is an auxiliary fermionic partner $\chi$ for each physical fermion $\psi$, of opposite chirality. It is used to write down a Wilson term $W$ which decouples the doublers. The auxiliary fermions interact with neither the gauge fields nor the Higgs field and decouple. This distinguishes them from the supplementary fermions in the mirror fermion model [7]. The action (6) has a global chiral invariance, with $\chi_{L(R)}$ transforming like $\psi_{R(L)}$.

The essential observation is that the auxiliary fields $\chi$ can be integrated out because they occur at most quadratically. This gives rise to an additional contribution

$$-\overline{\psi}_L W K^{-1} W \psi_L - \overline{\psi}_R W K^{-1} W \psi_R \qquad (7)$$

to the kinetic term of the $\psi$ field, leaving the rest of the action unchanged. The action is now of the Zaragoza form (4), with form factor given by

$$F_R^2 = \frac{s^2}{s^2 + w^2}, \qquad (8)$$

where $w(p) = \sum_\mu (1 - \cos p_\mu)$ is the Wilson term. Note that $F_R$ satisfies the requirements that it vanishes at the doubler momenta and that $F_R(0) = 1$. It is not differentiable at the doubler momenta, though.

For this form factor, the propagator (5) can be written as

$$S(p) = \frac{-i\slashed{s} + mF_R^2}{s^2 + w^2 + m^2 F_R^2}. \qquad (9)$$

There is no $w$-dependence in the numerator, contrary to the usual Wilson fermion propagator where such a term violates chiral symmetry and contributes via divergent loop integrals. This is in agreement with the conservation of chiral symmetry in the Roma approach.

## 4. ROMA PHASE DIAGRAM

Consider a chiral Yukawa model based on the Roma approach, in the formulation 'à la Zaragoza', with form factor $F_R$ (8). We can study the phase diagram in the mean-field approximation using techniques applied earlier for the Zaragoza model with the form factor (3), see refs. [6,5].

Here we focus on an $SU(2)_L \times SU(2)_R$ model in four dimensions, with $n_f$ fermion doublets and frozen modulus of the Higgs field. The phase diagram is specified in terms of the hopping parameter $k$ of the Higgs field and the Yukawa coupling $y > 0$.

In the small-$y$ region there is a paramagnetic phase at small $|k|$ and ferromagnetic (antiferromagnetic) phases at large positive (negative) values of $k$, as in the $O(4)$ model to which the system reduces for $y = 0$. These phases are separated by second order transition lines given by

$$\begin{aligned}
k_{\text{pf}}^{(W)}(y) &= \frac{1}{4} - y^2 \frac{n_f}{2} I_{\text{pf}}^{(W)}, \\
k_{\text{pa}}^{(W)}(y) &= -\frac{1}{4} - y^2 \frac{n_f}{2} I_{\text{pa}}^{(W)}, \qquad (10)
\end{aligned}$$

with

$$I_{\text{pf}}^{(W)} = \int_{-\pi}^{\pi} \frac{d^4p}{(2\pi)^4} \frac{F^4(p)}{s^2(p)},$$
$$I_{\text{pa}}^{(W)} = \int_{-\pi}^{\pi} \frac{d^4p}{(2\pi)^4} \frac{F^2(p)F_\pi^2(p)}{s^2(p)}, \quad (11)$$

where $F_\pi(p) \equiv F(p_1+\pi,\cdots,p_4+\pi)$. These curves meet in the point $A$ with coordinates

$$y_A = 1/\sqrt{n_f I_-^{(W)}}, \quad k_A = -I_+^{(W)}/4I_-^{(W)}, \quad (12)$$

where

$$I_\pm^{(W)} = I_{\text{pf}}^{(W)} \pm I_{\text{pa}}^{(W)}. \quad (13)$$

The phase structure at strong Yukawa coupling is similar. At $y = \infty$ we have the $O(4)$ model again, and there are PM, FM and AFM phases separated by the lines

$$k_{\text{pf}}^{(S)}(y) = \frac{1}{4} - \frac{1}{y^2}\frac{n_f}{2} I_{\text{pf}}^{(S)},$$
$$k_{\text{pa}}^{(S)}(y) = -\frac{1}{4} - \frac{1}{y^2}\frac{n_f}{2} I_{\text{pa}}^{(S)}, \quad (14)$$

where

$$I_{\text{pf}}^{(S)} = \int_{-\pi}^{\pi} \frac{d^4p}{(2\pi)^4} \frac{s^2(p)}{F^4(p)},$$
$$I_{\text{pa}}^{(S)} = \int_{-\pi}^{\pi} \frac{d^4p}{(2\pi)^4} \frac{s^2(p)}{F^2(p)F_\pi^2(p)}. \quad (15)$$

These lines meet in the point $B$ parametrized by

$$y_B = \sqrt{n_f I_-^{(S)}}, \quad k_B = -I_+^{(S)}/4I_-^{(S)}, \quad (16)$$

with $I_\pm^{(S)}$ defined analogously to eq. (13).

Note that the integrands of $I_{\text{pf,pa}}^{(S)}$ (15) are just the inverse of the integrands of $I_{\text{pf,pa}}^{(W)}$ (11). Nevertheless, one can show that for any form factor both $I_-^{(W)} > 0$ and $I_-^{(S)} > 0$, so that the points $A$ and $B$ are guaranteed to exist.

For the Roma form factor (8), one finds

$$I_{\text{pf}}^{(W)} = 2.5703\ 10^{-2}, \quad I_{\text{pf}}^{(S)} = 348.01,$$
$$I_{\text{pa}}^{(W)} = 7.3343\ 10^{-3}, \quad I_{\text{pa}}^{(S)} = 158.70. \quad (17)$$

The small and large-$y$ regions of the phase diagram for $n_f = 2$ are shown in fig. 1. We refer to ref. [5] for a discussion of the intermediate region.

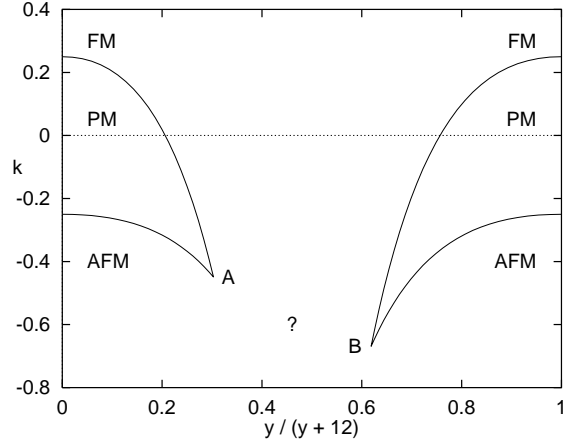

Figure 1. Partial phase diagram for the Roma $SU(2)_L \times SU(2)_R$ chiral Yukawa model with $n_f = 2$, in the mean-field approximation.

A special situation arises when the integrals $I_{\text{pf,pa}}^{(S)}$ diverge, as happens for the form factor of eq. (3). In this case the point $B$ lies on the line $y = \infty$, and the phase transition lines at large $y$ fall onto this line [5].

## REFERENCES

1. A. Borelli, L. Maiani, G.C. Rossi, R. Sisto and M. Testa, Nucl. Phys. B333 (1990) 335.
2. J.L. Alonso, Ph. Boucaud, J.L. Cortés and E. Rivas, Nucl. Phys. B (Proc. Suppl.) B17 (1990) 461; Mod. Phys. Lett. A5 (1990) 275; Phys. Rev. D44 (1991) 3258.
3. M.J. Dugan and L. Randall, Nucl. Phys. B382 (1992) 419.
4. J.L. Alonso, Ph. Boucaud, J.L. Cortés, F. Lesmes and E. Rivas, Nucl. Phys. B407 (1993) 373.
5. J.L. Alonso, Ph. Boucaud, F. Lesmes and A.J. van der Sijs, preprint DFTUZ/94/14, LPTHE Orsay-94/81.
6. J.L. Alonso, Ph. Boucaud, F. Lesmes and E. Rivas, Phys. Lett. 329B (1994) 75.
7. I. Montvay, Phys. Lett. 199B (1987) 89; 205 (1988) 315.